\documentclass{article}
\usepackage[a4paper, total={130mm, 230mm}]{geometry}
\usepackage{authblk} 
\usepackage{hyperref}

\newcommand\blfootnote[1]{%
  \begingroup
  \renewcommand\thefootnote{}\footnote{#1}%
  \addtocounter{footnote}{-1}%
  \endgroup
}

\usepackage{cite}
\usepackage{amsmath,amssymb,amsfonts}
\usepackage{algorithmic}
\usepackage{graphicx}
\usepackage{textcomp}
\usepackage{xcolor}
\usepackage{url}
\usepackage{array}  
\usepackage[colorinlistoftodos]{todonotes}
\usepackage[export]{adjustbox} 

\newcommand{\R}{\ensuremath{\mathbb{R}}}

\newcommand{\cY}{\mathcal{Y}}
\newcommand{\cX}{\mathcal{X}}

\newcommand{\cZ}{\mathcal{Z}}

\newcommand{\cD}{\mathcal{D}}
\newcommand{\cT}{\mathcal{T}}

\def\BibTeX{{\rm B\kern-.05em{\sc i\kern-.025em b}\kern-.08em
  T\kern-.1667em\lower.7ex\hbox{E}\kern-.125emX}}
  
\begin{document}

\title{\textbf{Improving Machine Hearing \\on Limited Data Sets}
\blfootnote{This work was supported by the Uni:docs Fellowship Programme for Doctoral Candidates in Vienna, by the Vienna Science and Technology Fund (WWTF) projects SALSA (MA14-018) and CHARMED (VRG12-009), by the International Mobility of Researchers (CZ.02.2.69/0.0/0.0/16 027/0008371), and by the project LO1401. Infrastructure of the SIX Center was used for computation.}
}
\author[1,2]{Pavol~Harar}
\author[1]{Roswitha~Bammer}
\author[1]{Anna~Breger}
\author[1]{\\Monika~D\"orfler}
\author[2]{Zdenek~Smekal}
\affil[1]{\footnotesize Faculty of Mathematics, NuHAG, University of Vienna, Vienna, Austria, \href{mailto:pavol.harar@univie.ac.at}{pavol.harar@univie.ac.at}}
\affil[2]{\footnotesize Department of Telecommunications, Brno University of Technology, Brno, Czech Republic}
\date{}




\maketitle

\begin{abstract}
Convolutional neural network~(CNN) architectures have originated and revolutionized machine learning for images. In order to take advantage of CNNs in predictive modeling with audio data, standard FFT-based signal processing methods are often applied to convert the raw audio waveforms into an image-like representations (e.g. spectrograms). Even though conventional images and spectrograms differ in their feature properties, this kind of pre-processing reduces the amount of training data necessary for successful training. In this contribution we investigate how input and target representations interplay with the amount of available training data in a~music information retrieval setting. We compare the standard mel-spectrogram inputs with a~newly proposed representation, called Mel scattering. Furthermore, we investigate the impact of additional target data representations by using an augmented target loss function which incorporates unused available information. We observe that all proposed methods outperform the standard mel-transform representation when using a~limited data set and discuss their strengths and limitations. The source code for reproducibility of our experiments as well as intermediate results and model checkpoints are available in an online repository.
\end{abstract}
\vspace{0.5em}



\section{Introduction}
Convolutional neural networks~(CNNs)~\cite{lecun1999object}, a~class of deep neural networks~(DNNs) architectures, originated in image processing and have revolutionized computer vision. The idea of CNNs is the introduction of locality and weight-sharing in the first layers of a~DNN, i.e. using convolutional layers. This leads to the extraction of local patterns, which are searched for over the entire image using the same filter kernels. By intermediate pooling operators, the extension of the local search increases across the layers and additionally introduces stability to local deformations,~\cite{mallat2016understanding}. 

Using the principles of CNNs in computer vision to solve problems in machine hearing, including music information retrieval~(MIR), has equally led to surprising successes in various applications. However, the data processing pipeline needs to be altered: the actual signal of interest, the raw audio signal, is not directly used as input to the network. Usually, it is first pre-processed into an image, allowing for a~time-frequency interpretation. Typical representations include the spectrogram or modifications thereof. This step leads to a~reduction of data needed for training~\cite{oriol}. 

In this paper we improve the performance of CNNs, which are trained with the standard mel-spectrogram (MT) \footnote{We abbreviate with MT, i.e. "mel-transform", in order not to collide with further abbreviations.} input representation and limited amount of training data. To do so, we propose an alternative input representation called \textit{Mel scattering}~(MS), which uses the main concept of \textit{Gabor scattering}~(GS), introduced in \cite{bammer2017gabor}, in combination with a~mel-filter bank. Moreover, we improve the learning results by transforming the target space within an \textit{augmented target loss function}~(AT), introduced in~\cite{breger2019orthogonal}.

The paper is organized as follows: In Section~\ref{Sec:LearnBas} we introduce the learning setup and the data used in the numerical experiments. In Section~\ref{Sec:TFR} we present the MT, and proceed to the definitions of GS and MS. AT is explained in Section~\ref{Sec:AT}. In Section~\ref{Sec:ExpRes} we compare the results of the proposed representations by evaluating the classification results of an instrumental sounds data set, serving as a~toy data set for experiments with different amount of training data.

\section{Learning from Data}\label{Sec:LearnBas}
Let $\mathcal{D}\subset \cX$ be a~data set in an 
input space $ \cX$, together with some information about the data, often called "annotation", which
is given in the target space and denoted by $\cT\subset \cY$. Learning the relationship between $\mathcal{D}$ and their annotations in $\cY$ can then 
be understood as looking for a~function $\psi: \cX\mapsto\cY$, which describes with sufficient accuracy the 
desired mapping. The accuracy is usually measured by a~loss function, which is optimized in each iteration step of the training process to update the weights. Once the learning process is finished, e.g. via a~stopping criterion, this results in a~parameter vector $\theta $ determining a~particular model within the previous determined architecture. 

Further, given a~hypothesis space parametrized by $\theta$, and a~
set of annotated data $\mathcal{Z}_m = \cD\times \cT = \{ (x_1, y_1 ) , \ldots, (x_m, y_m ) \}$, we learn a~model $\psi_\theta$. Let the estimated targets be denoted by $\hat{y}_i = \psi_\theta (x_i) $ and
define the empirical loss function $E_{\cZ_m}$ as 
\[E_{\cZ_m} (\psi_\theta) = \frac{1}{m} \sum_{i = 1}^m L( y_i , \hat{y}_i) . \]
Common, important examples of loss functions include the quadratic loss $ L( y_i , \hat{y}_i) = (\hat{y}_i- y_i ) ^2 $, 
and the categorical cross-entropy loss~(CE). The latter is the concatenation of the softmax function on the output 
vector $\hat{\mathbf{y}} = (\psi_\theta (x_1), \ldots, \psi_\theta (x_m))$ and the cross-entropy loss; in other words, in 
the case of categorical cross-entropy, we have 
$$ L( y_i , \hat{y}_i)  = - y_i \log\,\frac{e^{\hat{y}_i}}{\sum_{j = 1}^m e^{\hat{y}_j}}.$$

\subsection{Data Set used for Experiments}
\label{Data set}
For the classification experiments presented in Section~\ref{Sec:ExpRes}, the GoodSounds data set~\cite{romani2015real} is used. It contains monophonic recordings of single notes or scales played by different instruments. From each file, we have removed the silence with SoX v14.4.2 library\footnote{https://launchpad.net/ubuntu/+source/sox/14.4.1-5}. The output rate was set to $44.1$\,kHz with $16$\,bit precision. We have split each file into segments of the same duration ($1$\,s = $44\,100$\,samples) and applied a~Tukey window in order to smooth the onset and offset of the segment, thus preventing the undesired artifacts after applying the short-time Fourier transform (STFT).
Since the classes were not equally represented in the data set, we needed to introduce an equalization strategy. To avoid extensive equalization techniques, we have used only classes which spanned at least 10\% of the whole data set, namely clarinet, flute, trumpet, violin, alto saxophone and cello.
More precisely, during the process of cutting the audio samples into $1\,$s segments, we introduce increased overlap for instrument recordings with fewer samples, thus utilizing a~variable stride. This resulted in oversampling in underrepresented classes by overlapping the segments.

\section{Time-Frequency Representations of Audio} \label{Sec:TFR}
Classical audio pre-processing tools such as the mel-spectrogram rely on some localized, FFT-based analysis. 
The idea of the resulting time-frequency representation is to separate the variability in the signal with respect to time and frequency, respectively. 
However, for audio signals which are relevant to human perception, such as music or speech, significant variability happens on very different time-levels: the frequency content itself can be determined within a~few milliseconds. Variations in the amplitude of certain signal components, e.g. formants or harmonics, have a~much slower frequency and should be measured on the scale of up to few seconds. Longer-term musical developments, which allow, for example, to determine musical style or genre, happen on time-scales of more than several seconds. The basic idea of Gabor Scattering, as introduced in~\cite{bammer2017gabor}, see Section~\ref{Sec:GSMS}, is to capture the relevant variability at different time-scales and separate them in various layers of the representation. 

We first recall (mel-)spectrograms and turn to the definition of the scattering transforms in Section~\ref{Sec:GSMS}.

\subsection{Spectrograms and Mel-Spectrograms}
\label{Spec}

Standard time-frequency representations used in audio-processing 
are based on STFT. Since we are interested in obtaining several layers of time-frequency representations, we define STFT as {\it frame-coefficients} with respect to time-frequency-shifted versions of a~basic window. To this end, we introduce the following operators in some Hilbert space $\mathcal{H}$.
\begin{itemize}
\item The translation (time shift) operator 
 for a~function $f \in \mathcal{H}$ and $t \in \mathbb{R}$ is defined as $T_x f(t):=f(t-x)$ for all $x \in\mathbb{R}.$
\item The modulation (frequency shift) operator
for a~function $f \in \mathcal{H}$ and $t \in \mathbb{R}$ is defined as $M_\omega f(t):=e^{2\pi i t \omega}f(t) $ for all $\omega \in \mathbb{R}.$
\end{itemize}
Now the STFT $V_g f$ of a~function $f \in \mathcal{H}$ with respect to a~window function $g\in \mathcal{H}$ can be easily seen to be 
$V_g f(x,\omega)=\langle f,M_\omega T_x g \rangle $ with the corresponding spectrogram $|V_gf(x,\omega)|^2$. The set of functions
$$\mathit{G}(g,\alpha,\beta)=\{ M_{\beta j}T_{\alpha k}g : (\alpha k, \beta j)\in \Lambda\}$$ is a~the Gabor system and is called Gabor frame~\cite{grochenig2001foundations}, if there exist positive frame bounds $A, B>0$ such that for all $f \in \mathcal{H}$
\begin{equation}
A \| f\|^2 \leq \sum_{k}\sum_{j} |\langle f, M_{\beta j}T_{\alpha k}g\rangle |^2\leq B \| f\| ^2 .
\label{eq:Gabor}
\end{equation}
Subsampling $V_g f$ on a~separable lattice $\Lambda = \alpha \mathbb{Z}\times\beta\mathbb{Z}$ we obtain the frame-coefficients of $f$ w.r.t $\mathit{G}(g,\alpha,\beta)$. Choosing $\Lambda$
thus corresponds to picking a~particular hop size in time and a~finite number of frequency channels. 

The mel-spectrogram $MS_g(f)$ is defined as the result of weighted averaging $|V_g f(\alpha k,\beta j)|^2$:
$$MS_g(f)(\alpha k ,\nu) = \sum_j |V_gf(\alpha k,\beta j)|^2\cdot \Upsilon_\nu(j),$$
where $\Upsilon_\nu$ are the mel-filters for $\nu = 1,...,K$ with $K$ filters.

\subsection{Gabor Scattering and Mel Scattering}\label{Sec:GSMS}
We next introduce a~new feature extractor called Gabor scattering, inspired by Mallat's scattering transform~\cite{mallat} and first introduced in~\cite{bammer2017gabor}. In this contribution, we further extend the idea of Gabor-based scattering by adding a~mel-filtering step in the first layer. The resulting transform is called Mel scattering. Since the number of frequency channels is significantly reduced by applying the filter bank, the computation of MS is considerably faster.
GS is a~feature extractor for audio signals, obtained by an iterative application of Gabor transforms (GT), a~non-linearity in the form of a~modulus function and pooling by sub-sampling in each layer. Since most of the energy and information of an input signal is known to be captured in the first two layers, cp.~\cite{anden2014deep}, we only introduce and use the output of those first layers, while in principle scattering transforms allow for arbitrarily many layers. In~\cite{bammer2017gabor}, it was shown that the output of specific layers of GS lead to invariances w.r.t. certain signal properties. 

Coarsely speaking, the output of the first layer is invariant w.r.t. envelope changes and mainly captures the frequency content of the signal, while the second layer is invariant w.r.t. frequency and contains information about the envelope. For more details on GS and a~mathematical description of its invariances see~\cite{bammer2017gabor}.\\

In the following, since we deal with discrete, finite signals $f$, we let $\mathcal{H} =\mathbb{C}^\mathcal{L}$, where $\mathcal{L}$ is the signal length, and $f_\ell \in \mathbb{C}^{\mathcal{L}_{\ell}}$ for $\ell = 1,2$. The lattice parameters of the GT, i.e. $\Lambda_{\ell} = \alpha_{\ell} \mathbb{Z}\times\beta_{\ell}\mathbb{Z}$, can be chosen differently for each layer.

The first layer, which is basically a~GT, corresponds to 
\begin{align}
 f_1[\beta_1 j](k) = |\langle f, M_{\beta_1 j}T_{\alpha_1 k}g_1\rangle|,
 \label{eq:layer}
\end{align}
and the second layer can be written as
\begin{align}
 f_2[\beta_1 j,\beta_2 h](m) = |\langle f_{1}[\beta_1 j], M_{\beta_2 h}T_{\alpha_2 m}g_2\rangle|.
\end{align}
Note that the input function of the second layer is $f_1,$ where the next GT is applied separately to each frequency channel $\beta_1 j$.
To obtain the \textit{output} of one layer, one needs to apply an output generating atom $\phi_\ell$, cp.~\cite{bammer2017gabor}:

\begin{align} 
 f_{\ell} [\beta_1 s ,...,\beta_{\ell} j ] \ast \phi_{\ell}(k) = |\langle f_{\ell -1}, M_{\beta_{\ell} j}T_{\alpha_{\ell} k}g_1\rangle|\ast \phi_{\ell}, 
 \label{eq:output}
\end{align}
for $\ell \in \mathbb{N}$ in general and in our case $\ell = 1,2.$

The output of the feature extractor is the collection of these coefficients \eqref{eq:output} in one vector, which is used as input to a~machine learning task.
Based on the GS we want to introduce an additional mel-filtering step. The idea is to reduce the redundancy in spectrogram by frequency-averaging.
The expression in \eqref{eq:layer} is then replaced by
\begin{align}
  f_1[\nu](k)= \sum_j |\langle f_{0}, M_{\beta_1 j}T_{\alpha_1 k}g_1\rangle|\cdot \Upsilon_{\nu}(j),
  \label{eq:Mel}
\end{align}
where $\Upsilon_{\nu}$ corresponds to the mel-filters, as introduced in Section~\ref{Spec}.
The other steps of the scattering procedure remain the same as for GS, i.e. performing another GT to obtain layer $2$ and afterwards applying an output generating atom in order to obtain the MS coefficients.
The output of GS and MS can be visually explained by Figure~\ref{fig:vis}. 
The naming Output~A~displays either the output of Equation
\eqref{eq:layer} in case of GS or Equation \eqref{eq:Mel} in the MS case. The Output~B shows the spectrogram after applying the output generating atom and Output~C illustrates the output of the second layer.

\section{Augmented Target Loss Function}
\label{Sec:AT}
In the previous sections we introduced different input data representations for subsequent classification via deep learning. In the following we want to investigate possible enhancement with alternative output/target data representations. To do so, we use an augmented target loss function, a~general framework is introduced in~\cite{breger2019orthogonal}. It allows to integrate known characteristics of the target space via informed transformations on the output and target data. We now recall a~general formulation of AT from~\cite{breger2019orthogonal} and describe subsequently in detail, how it can be applied on the studied audio data.

Our training data is given by the MT of the sounds as inputs together with instrument classes as targets, introduced in Section~\ref{Data set}. The inputs to the network are thus arrays $\{x_i\}_{i=1}^m\subset \R^{120 \times 160}$ and have associated target values $\{y_i\}_{i=1}^m\subset \{0,1\}^6$, corresponding to the $6$ instrument classes. As described in Section~\ref{Sec:LearnBas}, in each optimization step for the parameters of the neural network, the network's output $\{\hat{y}_i\}_{i=1}^m\subset\R^6$ is compared with the targets $\{y_i\}_{i=1}^m$ via an underlying loss function $L$. 
However, training data often
naturally contains additional important 
target information that is not used in 
the original representation. We aim to incorporate such information 
tailored to the particular learning 
problem, enhancing the information content from the original target representation. 
Following the definition in \cite{breger2019orthogonal}, the augmented target loss function is given by

\begin{equation}\label{eq:erer}
\begin{split}
L_{AT} \big(y_i,\hat{y}_i \big) = \sum_{j=1}^n \lambda_j L_j \big(\mathfrak{T}_j(y_i),\mathfrak{T}_j(\hat{y}_i) \big). 
\end{split}
\end{equation}
Here, for all $j=1,\ldots,n$, we let $\lambda_j >0$ be an adjustable weight of $L_j$, which is some standard loss function and $\mathfrak{T}_j:\{0,1\}^6 \rightarrow \R^{d_j}$ is a~transformation which encodes the additional information on the target space.

Here, $\mathfrak{T}_1$ corresponds to the identity on $\mathbb{R}^6$, i.e. no transformation is applied in the first component, where $L_1$ is the categorical cross-entropy loss~\cite{zhang2018generalized}. For $j= 2, \dots , n$, we choose the dimension $d_j = 1$ and $L_j$ to be the mean squared error. The incorporation of additional information on the GoodSounds data set is described in detail in the following section.

\subsection{Design of Transformations}\label{Sec:Design}
We heuristically choose $d = 16$ transformations $\mathfrak{T}_2,\dots, \mathfrak{T}_{17}$ that use target characteristics (features) arising directly from the particular target class, with $\mathfrak{T}_j: \{0,1\}^6 \to \R$, for $j = 2, \dots, 17$. Amongst others the features are chosen from the enhanced scheme of taxonomy~\cite{von1961classification} and from the table of frequencies, harmonics and under tones~\cite{FrequencyRanges}. We choose transformations that provide information that is naturally contained in the underlying instrument classes. The additional terms in the loss function \eqref{eq:erer} shall enable to penalize common classification errors. In this experiment, the transformations are given by the inner product of the output/target and the feature vector. E.g. we directly know to which instrument family an instrument belongs and distinguish between woodwind, brass and bowed instruments, and moreover between chordophone and aerophone instruments. Let's assume a~target vector $y_i (j) = \delta_{ij}$, corresponds, respectively, to the instruments clarinet, flute, trumpet, violin, saxophone and cello, and the output of the network is $\hat{y}_i = (a_1, a_2, a_3, a_4, a_5, a_6)\in \R^6$. The feature vector $v_1 = (1,1,0,0,1,0)$ then captures the information "target instrument is from family woodwind". The transformation may be defined by $\mathfrak{T}_1(y_i) = \langle y_i, v_1 \rangle$ in order to incorporate this information. Additionally, by choosing $\lambda_j$, we can weight the amount of penalization for wrong assignments in $(\mathfrak{T}_1(y_i) - \mathfrak{T}_1(\hat{y}_i))^2$. Amongst others we also use minimum and maximum frequencies of the instrument as features. E.g. the feature corresponding to minimum frequency $v_2 = (b_1, b_2, b_3, b_4, b_5, b_6)\in \R^6$. Again the transformation is given by $\mathfrak{T}_2(y_i) = \langle y_i, v_2 \rangle$. Choosing the right penalty for this feature could prohibit that instruments belonging to the same instrument family are classified wrong, e.g. a~cello that would be classified as a~violin. One can think about AT as a~method to more precisely define the measure of distance between the predicted and target classes.

\section{Numerical Experiments}
\label{Sec:ExpRes}

In the numerical experiments, we compare the performances of CNNs trained using the CC loss and time-frequency representations mentioned in Section~\ref{Sec:TFR}. As a~baseline, we use the results of MT. Furthermore we compare the baseline with the results of MT with AT loss as introduced in Section~\ref{Sec:AT}. The overall task is a~multi-class classification of musical instruments based on the audio signals introduced in Section~\ref{Data set}.

\subsection{Computation of Signal Representations}
The raw audio signals were transformed into MT, MS and GS time-frequency representations, using the Gabor-scattering~v0.0.4 library~\cite{gabor-scattering}. The library contains our Python implementation of all previously mentioned signal representations, with the aim to provide the community with an easy access to all of the transformations. The library's core algorithms are based on Scipy~v1.2.1~\cite{scipy, oppenheim1999discrete, griffin1984signal} implementation of STFT and mel-filter banks from Librosa~v0.6.2 library~\cite{librosa0.6.2}. 

\begin{figure}[t]
\begin{center}
\includegraphics[width=\linewidth, max height=13cm]{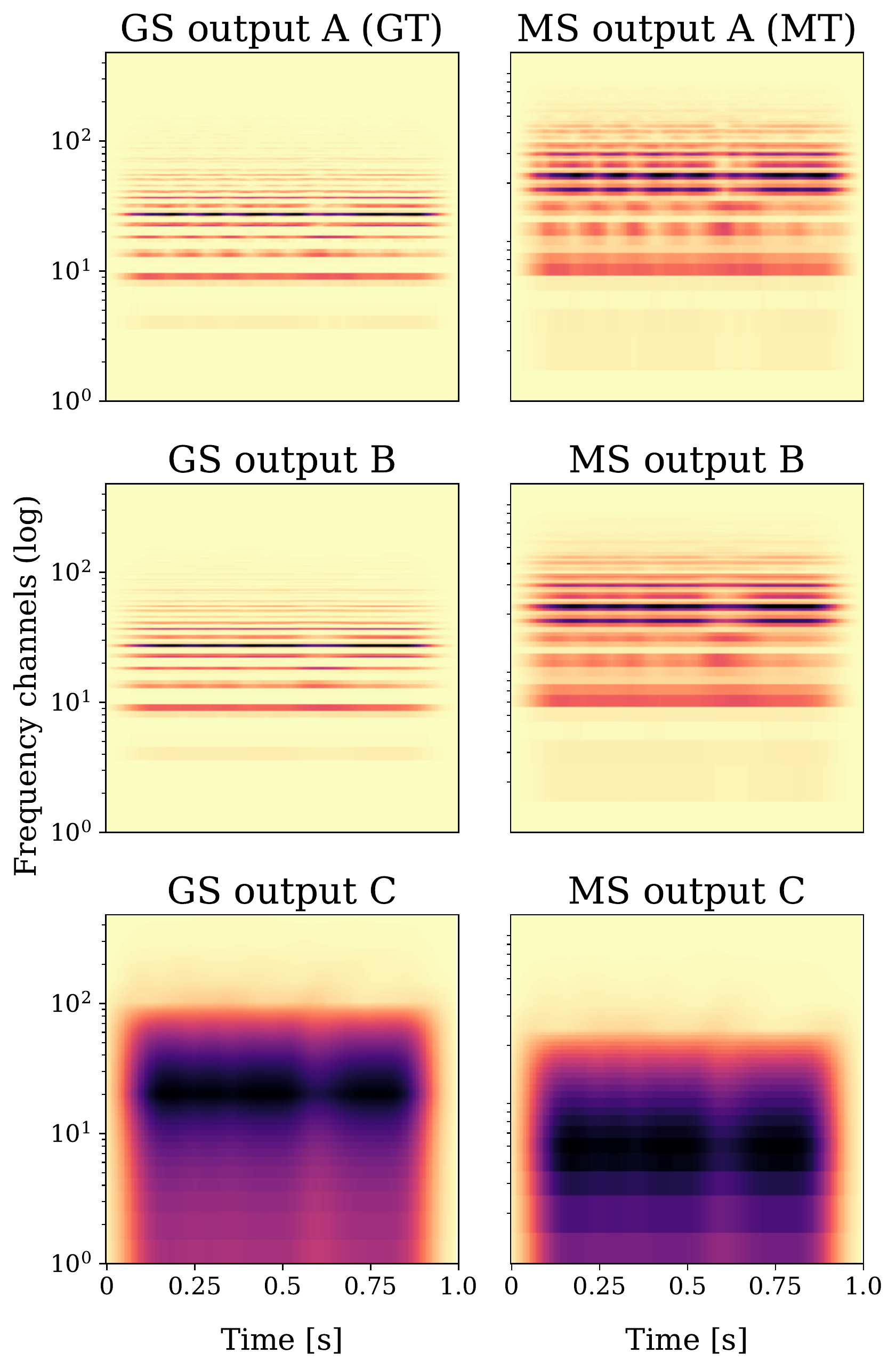}
\caption{Visualization of time-frequency transformations.}
\label{fig:vis}
\end{center}
\end{figure}

All the representations are derived from GT. In order to have a~good resolution in time and frequency for our classification task, we have chosen the parameters heuristically. The final shapes of the representations are shown in Table~\ref{table:shapes}. The three dimensional output of GS contains the GT and outputs of layer 1 and 2 of the GS cf.\cite{bammer2017gabor}, the same applies to MS. The visualizations of the time-frequency transformations of an~arbitrary training sample are shown in Figure~\ref{fig:vis}.

\subsection{Deep Convolutional Neural Network}
We implemented our experiment in Python 3.6. A~CNN was created and trained from scratch on Nvidia~GTX~1080~Ti GPU in Keras~2.2.4 framework~\cite{chollet2015keras} using the described training set split into batches of size $128$. We used an architecture consisting of four convolutional stacks. Each of them consists of a~convolutional layer, rectified-linear unit activation function and average pooling. These stacks were followed by a~fully connected layer with softmax activation function. Each network had to be adjusted slightly, because the input shapes changed according to the time-frequency representation used (GS has 3 channels, MT has less frequency channels etc.). We have tried to make the results as comparable as possible, therefore the networks differ only in the number of channels of the input layer, the rest of the network is only affected by the number of frequency channels, which thanks to pooling did not cause significant difference in the number of trainable parameters. All networks have comparable number of trainable parameters within the range from $81\,042$ to $83\,882$. The weights were optimized using Adam optimizer~\cite{adam}. Reproducible open source code can be found in the repository~\cite{gs-ms-mt}.

\begin{figure}
\begin{center}
\includegraphics[width=\linewidth,  max height=20cm]{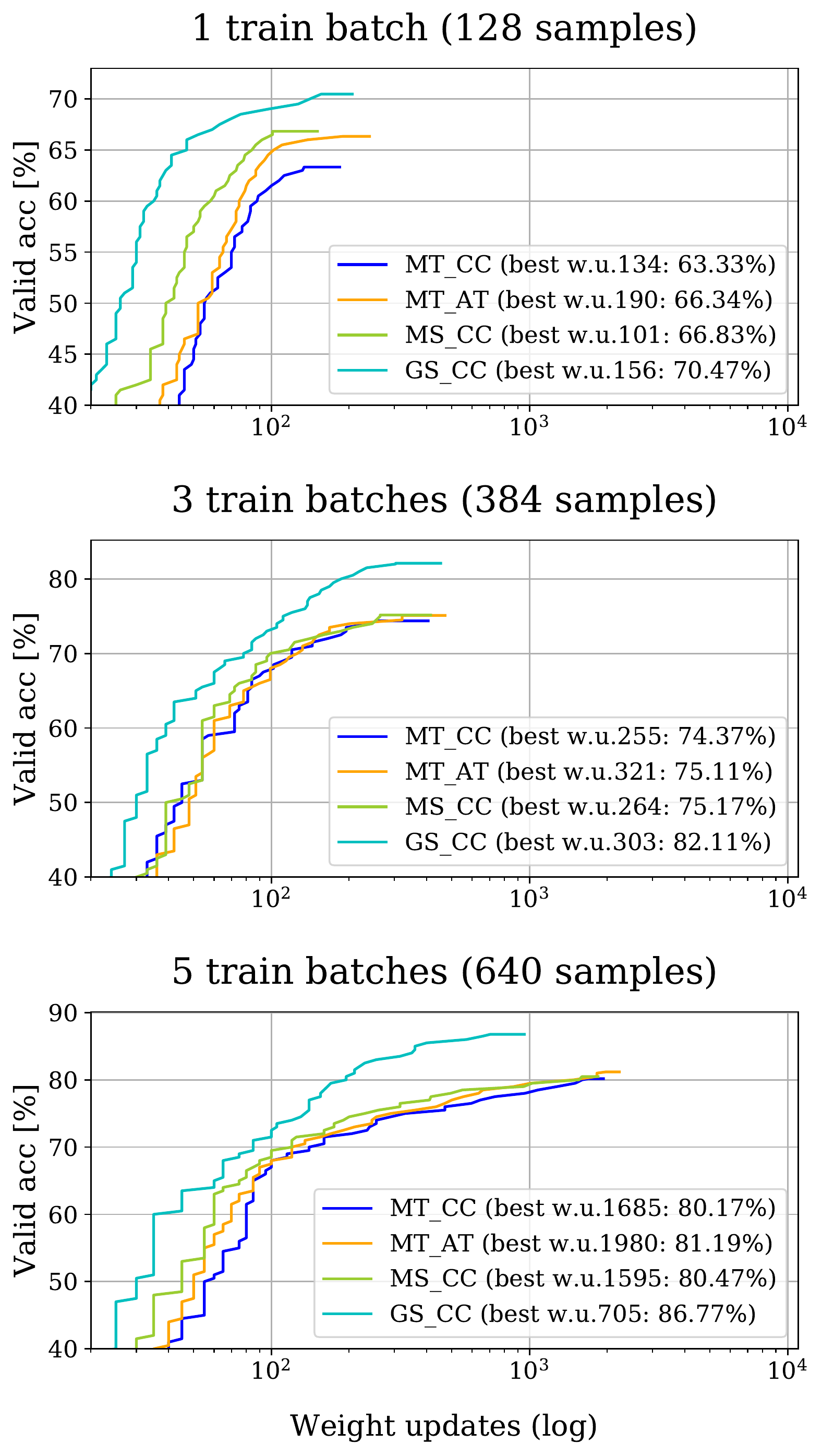}
\caption{CNN performance milestone reached over number of weight updates. 
The computational effort in all experiments was limited to 11\,000 weight updates.
Figure notation: Valid acc\,--\,Accuracy performance metric measured on the validation set,
Best w.u.\,--\,Weight update after which the highest performance was reached.}
\label{fig:performance}
\end{center}
\end{figure}

\subsection{Training and Results}
All the samples were split into training, validation and testing sets in such a~way that validation and testing sets have exactly the same number of samples from each class, while this holds for training set only approximately. Segments from audio files that were used in validation or testing were not used in training to prevent leaking of information. Detailed information about the used data, stride settings for each class, obtained number of segments and their split can be found in the repository~\cite{gs-ms-mt}.

In total we have trained 36 different models (MT, MS, GS with CC and MT with AT trained on 9~training set sizes), with the following hyper-parameters: number of convolutional kernels in the first 3 convolutional layers is 64 each, learning rate is 0.001, $\lambda$ of AT is 10 and $\lambda$ of $L_2$ weight regularization is 0.001. As a~baseline we have used MT with a~standard CC loss function as implemented in the Keras framework and described in detail in Section~\ref{Sec:LearnBas}. The computational effort was limited to 11\,000 weight updates. Time necessary for one weight update of each model is shown in Table~\ref{table:shapes}.

\begin{figure}
\begin{center}
\includegraphics[width=\linewidth,  max height=16cm]{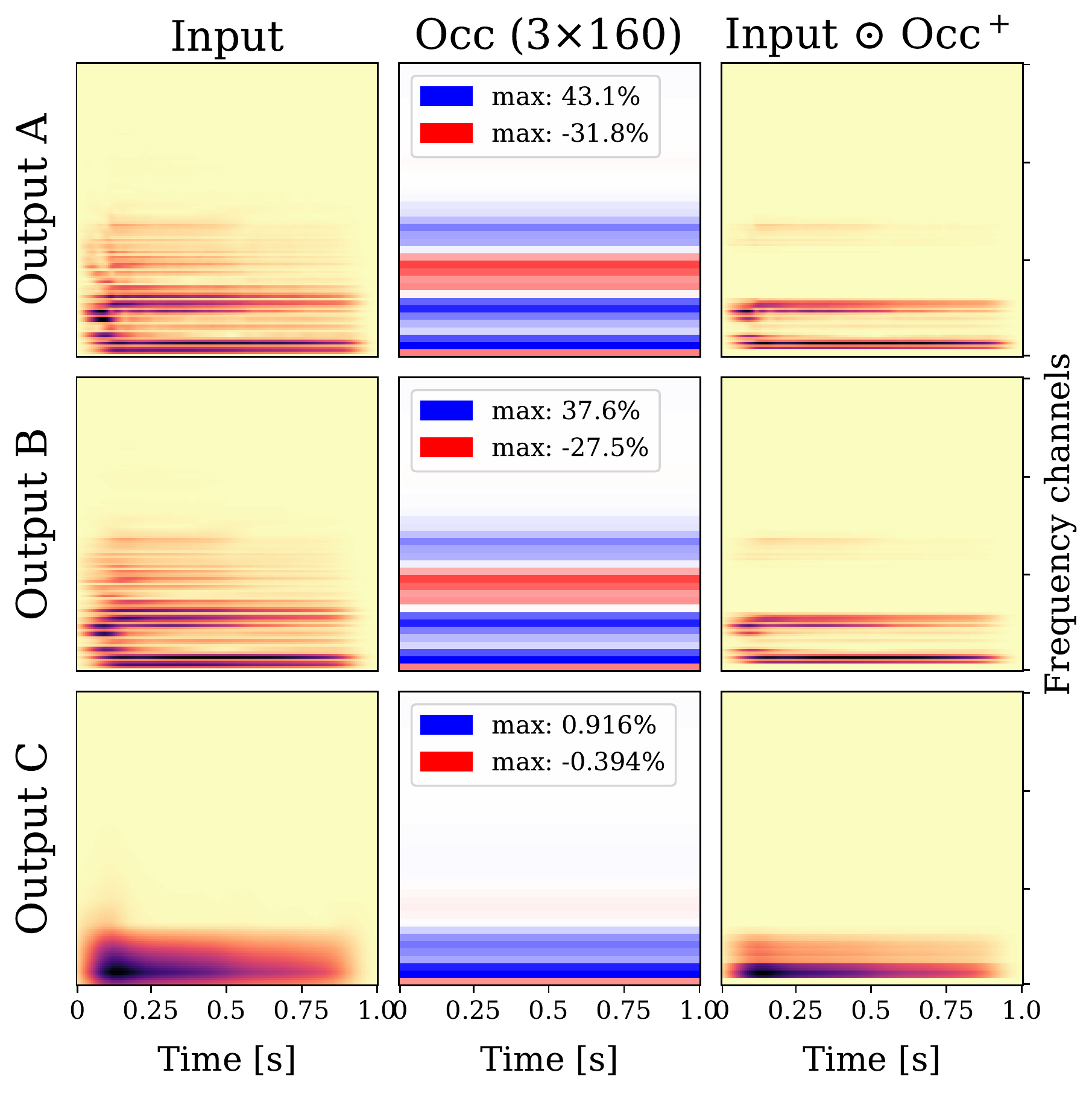}
\caption{Visualization of occlusion maps and frequency channel importance based on the best performing model trained on 1 batch of MS. Signal shown is randomly selected alto sax sample.
Figure notation: Input\,--\,input representation for CNN.
Occ\,--\,occlusion map created by sliding occlusion window.
Input~$\odot$~Occ$^+$\,--\,Elementwise multiplication of input with positive semidefinite occ (negative elements were changed to zeros before multiplication).
Blue and red colors\,--\,Positive and negative influence of particular frequency channel bin on the model performance.}
\label{fig:occlusion}
\end{center}
\end{figure}

Table~\ref{table:1} shows the highest achieved accuracies of the CNN models trained with MT for different training set sizes along with the improvements of this baseline by proposed methods. Accuracy is computed as a~fraction of correct predictions to all predictions. In Figure~\ref{fig:performance} we compare the number of weight updates necessary to surpass a~certain accuracy threshold for all proposed methods. Occlusion maps~\cite{zeiler2014visualizing} for a~random MS sample are visualized per 3~frequency bins in Figure~\ref{fig:occlusion}.

\begin{table}
\caption{Shapes and execution time}
\label{table:shapes}
\begin{center}
\begin{tabular}{| c | c | c | c |}
\hline															
TF  &  shape & CC & AT \\
\hline	
\hline	
GS & $3 \times 480 \times 160$ & 950\,ms & -     \\ 
MT & $1 \times 120 \times 160$ & 250\,ms & 320\,ms \\ 
MS & $3 \times 120 \times 160$ & 450\,ms & -     \\ 
\hline
\end{tabular}
\end{center}
Table notation: 
TF\,--\,Time-frequency representation. CC/AT\,--\,The execution time of one weight update during training with CC/AT loss function.
\end{table}

\begin{table}
\caption{Improvements of the MT Baseline}
\label{table:1}
\begin{center}

\begin{tabular}{| r | r | r | r | r |}
\hline
\multicolumn{5}{|c|}{Highest validation set accuracies}\\
\hline									
\hline									
NB	&	MT	&	MT$_\mathrm{AT}$	&	MS	&	GS	\\
\hline									
\bfseries1	& \bfseries63.33\,\%  &	\bfseries+3.01\,\%  & \bfseries+3.50\,\%   & \bfseries+7.15\,\%	\\
\bfseries3	& \bfseries74.37\,\%  &	\bfseries+0.74\,\%  & \bfseries+0.80\,\%   & \bfseries+7.74\,\%	\\
\bfseries5	& \bfseries80.17\,\%  &	\bfseries+1.02\,\%  & \bfseries+0.31\,\%   & \bfseries+6.60\,\%	\\
7	& 82.93\,\%  &	-1.12\,\%  & -0.09\,\%   & +5.63\,\%	\\
9	& 85.40\,\%  &	+0.95\,\%  & -0.43\,\%   & +5.28\,\%	\\
11	& 86.53\,\%  &	+0.33\,\%  & +1.26\,\%   & +5.57\,\%	\\
55	& 96.06\,\%  &	-0.27\,\%  & -0.27\,\%   & +2.52\,\%	\\
110	& 96.31\,\%  &	-0.04\,\%  & +0.06\,\%   & +2.53\,\%	\\
550	& 96.00\,\%  &	+0.74\,\%  & +0.48\,\%   & +3.12\,\%	\\
\hline
\end{tabular}

\vspace{1.5em}

\begin{tabular}{| r | r | r | r | r |}
\hline
\multicolumn{5}{|c|}{Corresponding testing set accuracies}\\
\hline									
\hline									
NB	&	MT	&	MT$_\mathrm{AT}$	&	MS	&	GS	\\
\hline									
\bfseries1	& \bfseries64.28\,\%  & \bfseries+2.73\,\%  & \bfseries+3.36\,\%  &	\bfseries+6.93\,\% \\
\bfseries3	& \bfseries75.61\,\%  & \bfseries+0.58\,\%  & \bfseries+0.32\,\%  &	\bfseries+7.26\,\% \\
\bfseries5	& \bfseries80.69\,\%  & \bfseries+0.79\,\%  & \bfseries+0.07\,\%  &	\bfseries+6.93\,\% \\
7	& 83.48\,\%  & -1.13\,\%  & +0.37\,\%  &	+6.30\,\% \\
9	& 86.30\,\%  & +0.54\,\%  & -0.43\,\%  &	+5.23\,\% \\
11	& 87.41\,\%  & -0.43\,\%  & +1.30\,\%  &	+4.85\,\% \\
55	& 96.27\,\%  & -0.20\,\%  & -0.31\,\%  &	+2.26\,\% \\
110	& 96.80\,\%  & -0.55\,\%  & -0.12\,\%  &	+2.21\,\% \\
550	& 96.72\,\%  & +0.27\,\%  & +0.07\,\%  &	+2.29\,\% \\
\hline
\end{tabular}

\vspace{0.5em}
\end{center}

Table notation: 
NB\,--\,Number of training batches with 128 samples each. MT, MS and GS\,--\,mel-spectrogram, Mel scattering and Gabor scattering as input representations with CC. MT here servers as a~baseline for comparison with other methods. MT$_\mathrm{AT}$\,--\, mel-spectrogram as input representation with AT. Testing set accuracies were evaluated after the epoch where the validation accuracy was the highest. 
\end{table}

\section{Discussion and Conclusions}

Our previous work on Gabor scattering showed that signal variability w.r.t. different time scales is separated by this transform, cf.~\cite{bammer2017gabor}, which is a~beneficial property for learning. The common choice of a~time-frequency representation of audio signals in predictive modeling is mel-spectrogram; hence, as a~natural step, we introduced MS in this paper, a~new feature extractor which combines the properties of GS with mel-filter averaging. We also investigated the impact of additional information about the target space through AT on the performance of the trained CNN.

From the results on GoodSounds dataset shown in Table~\ref{table:1}, we see that all proposed methods outperform the baseline (mel-spectrogram with categorical cross-entropy loss) on the first three most limited training sets, i.e. the data sets with the least amount of data. All proposed methods also show a~trend to achieve better results earlier in the training, as visible in Figure~\ref{fig:performance}. This trend seems to diminish with bigger training set sizes. Improvements on the last, biggest training set can be justified by the fact that this experiment was interrupted before it had the time to converge, therefore highlighting earlier successes of the proposed methods. 
From the newly proposed methods, AT is the least expensive in terms of training time, but on the other hand yields the smallest improvement in this experimental setup. Nevertheless, it has another advantage: it steers the training towards learning the penalized characteristics, e,g. to learn the characteristic of an instrument being or not being a~wood instrument if the information about this grouping is provided through AT. We believe that the positive effect of AT in this setup becomes obsolete with higher number of training batches because after training above a~certain accuracy threshold, the network already predicts the correct groups of classes and therefore can not gain from AT anymore. 

MS performed better than both MT and MT$_\mathrm{AT}$ for slightly higher cost of computation and also achieved the same performances earlier. GS outperformed all of the tested methods and showed an improvement over all training set sizes, however this might also suggest that GT (without mel-filtering) would be a~better input data representation for this task in the first place. As in GS, MS comprises exclusively the information of its MT origin. The separation of the embedded information into three distinct channels might be the reason for its success. The evidence is visible in Figure~\ref{fig:performance}, which shows MS reaching higher accuracies after less weight updates than MT, suggesting that the network did not have to learn similar separation during training. Also, the visualization in Figure~\ref{fig:occlusion} supports this by showing a~positive influence of Outputs~A and~B on the model's performance.

It remains to be said, that improvements which can be gained by using AT, MS or GS highly depend on the task being solved, on the choice of transformations based on the amount of additional available information for AT and on the correctly chosen parameters of the time-frequency representations. 

From what was stated above, we can conclude that AT provides a~more precise measure of distance between outputs and targets. That's why it can help in scenarios where the training set is not large enough to allow the learning of all characteristics, but can be penalized by AT. We suggest to use/experiment with the proposed methods for other data sets if there is not a~sufficient amount of data available or/and there exist reasonable transformations in the target space relevant to the task being solved. All proposed methods might be found useful also in scenarios with limited resources for training.

In order to obtain reliable statistical results on the various methods, it would be necessary to run all experiments several hundred times with different seeds. For the current contribution, such a~procedure was not included due to the restriction of computational resources and is thus left for future work.

\bibliographystyle{abbrv}
\bibliography{main}
\end{document}